\documentclass[aps,prd,superscriptaddress,nofootinbib,tighten,preprint]{revtex4}
\usepackage[utf8]{inputenc}
\usepackage[english]{babel}
\usepackage{amsmath}
\usepackage{subfigure}
\usepackage{amsfonts}
\usepackage{amssymb}
\usepackage{graphicx}
\newcommand{\be}{\begin{equation}}
\newcommand{\ee}{\end{equation}}
\newcommand{\bea}{\begin{eqnarray}}
\newcommand{\eea}{\end{eqnarray}}

\newcommand{\nn}{\nonumber}

\catcode`@=12

\begin{document}

\title{Explaining Xenon-1T signal with FIMP dark matter and neutrino mass in a $U(1)_{X}$ extension}

\author{Sarif Khan}
\email{sarif.khan@uni-goettingen.de}
\affiliation{    Institut f\"{u}r Theoretische Physik, Georg-August-Universit\"{a}t G\"{o}ttingen,
Friedrich-Hund-Platz 1, 37077 G\"{o}ttingen, Germany}

\begin{abstract} 
In the present work, we have extended the standard model by an abelian $U(1)_{X}$
gauge group and additional particles. In particular, we have extended the particle content
by three right handed neutrinos, two singlet scalars and two vector like leptons. Charged assignments under different
gauge groups are such that the model is gauge anomaly free and the anomaly contributions cancel among generations. 
Once the symmetry gets broken then three
physical Higgses are produced, one axion like particle (ALP), which also acts as the 
keV scale FIMP dark matter, is produced and the remaining component is absorbed by the extra gauge boson. 
Firstly, we have successfully generated neutrino mass by the type-I seesaw mechanism
for normal hierarchy with the $3\sigma$ bound on the oscillation parameters.
The ALP in the present model can explain the 
Xenon-1T electron recoil signal at keV scale through its coupling with the electron. 
We also have vector like leptons which help
in producing the dark matter from their decay by the freeze in mechanism. 
Electron and tauon get mass from dimensional-5 operators at Planck scale and if 
we consider the vevs $v_{1,2} \simeq 10^{12}$ GeV
then we can obtain the correct value of the electron mass but not the tauon mass. Vector like 
leptons help in getting the correct value of the tauon mass through 
another higher dimensional operator which also has a role in DM production by the 
$2 \rightarrow 2$ process, giving
the correct ballpark value of relic density for suitable reheat temperature of the Universe.
We have shown that the ALP production by the higher dimensional operator can explain
the electron, tauon mass and Xenon-1T signal simultaneously whereas the decay production can not 
explain all of them together.
\end{abstract}

\maketitle

\section{Introduction}
The standard model (SM) is a very successful theory in describing nature without any doubt.
Although its tremendous success, the SM has few flaws which can not be addressed within its particle content and gauge structure. The most noticeable limitations of the SM are the absence of 
a suitable dark matter (DM) candidate and neutrino masses.
The presence of DM is a very well established phenomenon and has been confirmed by many experiments
namely Galaxy rotation curve \cite{Sofue:2000jx}, 
Bullet cluster \cite{Clowe:2003tk, Harvey:2015hha}, gravitational lensing \cite{Bartelmann:1999yn}
and the measurements of the Cosmic Microwave Background (CMB) \cite{Hinshaw:2012aka,Ade:2015xua}.
The satellite borne CMB experiments, 
WMAP \cite{Hinshaw:2012aka} and Planck \cite{Ade:2015xua} have measured the DM relic density 
($\Omega h^2$)
with an unprecedented accuracy which is,
\begin{eqnarray}
0.1172 \leq \Omega h^2 \leq 0.1226\,.
\end{eqnarray}
Moreover, in the SM neutrinos are massless, 
but from oscillation experiments, it is well established that
neutrinos are massive in order to explain the flavour oscillation among 
the different flavours \cite{Cowan:1992xc, Fukuda:1998mi, Ahmad:2002jz, Eguchi:2002dm, An:2015nua, RENO:2015ksa, Abe:2014bwa, Abe:2015awa, Salzgeber:2015gua, Adamson:2016tbq, Adamson:2016xxw}. 
After a rigorous search of DM at the collider, direct detection and indirect detection 
experiments, finally, Xenon-1T collaboration has announced their discovery in
the searches of new physics with low-energy electronic recoil data. They
observed excess events over the known backgrounds in 1 to 7 keV range \cite{Aprile:2020tmw}.
The Xenon-1T experiment consisting of 1042 kg of cylindrical fiducial volume with 226.9 active
days {i.e.} in total 0.65 tonne-year exposure, has observed 285 electron recoil events
 in the range $1-7$ keV in compared to the expected $232 \pm 15$ background events,
which gives $3.3 \sigma$ Poissonian fluctuation. Xenon-1T considered many background models
as listed in \cite{Aprile:2020tmw}, among them, the prominent ones are the
Pb, Kr, Xe, I and solar neutrinos in the region of interest (ROI)
$1-210$ keV. With the 0.65 tonne-year exposure of SR1 they can not explain the excess events.
There are a couple of possible scenarios discussed in \cite{Aprile:2020tmw} which can explain the 
above excess namely solar axion,
neutrino with the magnetic moment and bosonic dark matter. The background model is rejected at the 
$3.5\, \sigma$, $3.2\,\sigma$ and $3\,\sigma$ for the solar axion, neutrino magnetic moment
and bosonic dark matter model, respectively.
The parameter space needed to explain Xenon-1T signal for solar axion           
and neutrino magnetic moment model is in strong tension with the stellar cooling 
\cite{Bertolami:2014wua, Ayala:2014pea, Viaux:2013lha, Giannotti:2017hny, DiLuzio:2020wdo}
and white dwarf \cite{Corsico:2014mpa}, globular cluster \cite{Diaz:2019kim}, respectively.
Moreover, if one considers the tritium background which is neither confirmed nor
excluded, then the significance of solar axion model is reduced to $2.1\,\sigma$
and the neutrino magnetic moment model to $0.9\,\sigma$.

In the present work, we are going to explain the signal by an axion like particle (ALP), $a$, 
which will be pseudo Nambu Goldstone boson (PNGB) produced due to a 
Peccei-Quinn type global symmetry breaking.
The general structure of the Lagrangian of this kind of ALP is,
\begin{eqnarray}
\mathcal{L}_{ALP} &=& (\partial_{\mu} a)^2 - m^{2}_{a} a^2 + 
\frac{g_{a\gamma\gamma}}{4} a F_{\mu\nu} \tilde{F}^{\mu\nu} 
+ \frac{g_{aee}}{2 m_{e}} (\partial_{\mu} a) \bar{e} \gamma^{\mu} \gamma_{5} e
\end{eqnarray}
By using the relation $\partial_{\mu}(\bar{e} \gamma^{\mu} \gamma_{5}e) = 
2 i m_{e} \bar{e} \gamma_{5} e$, we can write the above equation in the following way,
\begin{eqnarray}
\mathcal{L}_{ALP} &=& (\partial_{\mu} a)^2 - m^{2}_{a} a^2 + 
\frac{g_{a\gamma\gamma}}{4} a F_{\mu\nu} \tilde{F}^{\mu\nu} 
+ i g_{aee} a \bar{e} \gamma_{5} e
\label{ALP-lagrangian-general}
\end{eqnarray} 
As shown in \cite{Bloch:2020uzh}, the best fit values for explaining the Xenon-1T signal 
by the ALP are the following,
\begin{eqnarray}
m_{a} = 2.5\,\,{\rm keV}\,, g_{aee} = 2.5 \times 10^{-14}\,. 
\end{eqnarray}
To put the constraints on the dimensionless parameters, we can redefine the
couplings in the following way,
\begin{eqnarray}
g_{a\gamma\gamma} = \frac{\alpha_{em}}{2 \pi V} g^{a\gamma\gamma}_{eff},\,\,\,
g_{aee} = \frac{m_{e}}{V} g^{aee}_{eff}
\end{eqnarray}
where $V$ is the ALP decay constant, $g^{a\gamma\gamma}_{eff}$ and $g^{aee}_{eff}$  
are the effective coupling of axion to $\gamma \gamma$ and $ee$, respectively.
We can put constraint on the $g^{a\gamma\gamma}_{eff}$ parameter from the
cosmic X-ray background (CXB) for keV scale DM candidate (which corresponds to the 
frequency $2.98 \times 10^{17}$ Hz) as follows \cite{Hill:2018trh},
\begin{eqnarray}
\frac{g^{a\gamma\gamma}_{eff}}{g^{aee}_{eff}} \leq 3.1 \times 10^{-3} 
\left( \frac{2.5\,\,{\rm keV}}{m_{a}} \right)^{3/2} \times 
\left( \frac{2.7 \times 10^{-14}}{g_{aee}} \right)\,.
\end{eqnarray}   
In the literature, there are extensive studies to explain the Xenon-1T signal
by considering DM as its origin and can be found in the Refs.\,(\cite{Takahashi:2020bpq, Kannike:2020agf, Alonso-Alvarez:2020cdv, Fornal:2020npv, Boehm:2020ltd, Harigaya:2020ckz, Su:2020zny, DiLuzio:2020jjp, Chen:2020gcl, Bell:2020bes, Dey:2020sai, newBuch, Choi:2020udy, AristizabalSierra:2020edu, Paz:2020pbc, Lindner:2020kko, Budnik:2020nwz, Zioutas:2020cul, DelleRose:2020pbh, Dessert:2020vxy, Coloma:2020voz, Chao:2020yro, Cacciapaglia:2020kbf, Ko:2020gdg, Alhazmi:2020fju, Baek:2020owl, Li:2020naa, Inan:2020kif, Benakli:2020vng, Okada:2020evk, Choi:2020kch, Davoudiasl:2020ypv, He:2020wjs, Athron:2020maw, Anastasopoulos:2020gbu, Arias-Aragon:2020qtn, Choudhury:2020xui, Arcadi:2020zni, Takahashi:2020uio, Cao:2020oxq, 1808116, 1808091}). 
To accomplish the dark matter, neutrino mass and the general ALP model described above 
(as shown in Eq.\,(\ref{ALP-lagrangian-general})),
we have extended SM by additional particles and gauge group.  
In particular, we have extended the SM gauge structure by an additional
local gauge group $U(1)_{X}$, three right handed neutrinos, two singlet
scalars and two vector like leptons doublet. We have assigned the gauge charges to all the particles 
(SM as well as beyond SM particles) in such a way so that the gauge anomaly cancels automatically.
For the Higgs doublet among four degrees of freedom (d.o.f), three of them are absorbed by the $W^{\pm}$
and $Z$ bosons. The remaining four d.o.f for the two singlet scalars, one of them absorbed
by the extra gauge boson present due to the additional gauge group, one of them act as the axion like particle (ALP) and the remaining two become physical Higgses. Since the main motive apart from explaining the neutrino mass is to explain the Xenon-1T signal from DM point of view,
 in the present work, ALP (denoted as $a$) which is also feebly interacting massive particle (FIMP) 
dark matter takes a significant role in the phenomenology. Physical Higgses (defined as $h$, $h_1$ and $h_2$) do not actively take part in the
phenomenology we are interested in this work. Nevertheless if one studies the collider signature
of the vector like lepton then the physical Higgses play an important role in giving novel
signatures to detect vector like leptons at the collider. The charges of the two singlet Higgses are assigned in such a 
way that the ALP is massless and it can achieve mass from the higher dimensional operator
at the Planck scale. We consider a global symmetry, $U(1)_{g}$, like Peccei-Quinn symmetry and ALP is produced when the global symmetry gets broken at the intermediate
scale and the extra singlet scalars take vevs. The ALP coupling to two electrons also
appears when the global symmetry breaks and it is necessary to explain the Xenon-1T 
signal of electron recoil at keV range.  
We assign the global charges to the particles
in such a way so that $U(1)_{g} \times U(1)_{em} \times U(1)_{em}$ is anomaly free.
Because of this anomaly cancellation, the ALP coupling to two photons is 
suppressed and can evade the
present day strong bound from CXB \cite{Hill:2018trh}. 
Although in the present work, we are not discussing collider signature of the 
vector like lepton, still
it takes an important role in the production of the ALP by the freeze-in mechanism.
We can produce the ALP by making the associated Yukawa coupling strength
of ALP with the vector like lepton very small which is $\mathcal{O}(10^{-8})$.
 Moreover, the vector like 
lepton is heavy compared to the associated particles and we can have dimension-5 
 operator after eliminating the vector like lepton,
which can also produce DM by the $2 \rightarrow 2$ process for suitable values of the 
reheat temperature \cite{McDonald:2001vt, Hall:2009bx}.                 

Rest of the paper is organised in the following way. In Section \ref{model}, we have discussed the present model in detail. Neutrino mass and allowed parameter space after satisfying oscillation
have been covered in Section \ref{neutrino-mass}.
Section \ref{xenon-1T-discussion} focuses on the Xenon-1T signal and production of dark matter
by the freeze in mechanism. Finally we conclude in Section \ref{conclusion}.

\section{Model}
\label{model}
In the present work, we have considered a $U(1)_{X}$ extension of the SM gauge group.
Besides, the gauge group, SM particle has also been extended by three right handed neutrino, 
one pair of vector like leptons and two singlet scalars. All the particles are charged under
the extra gauge group and in particular, they are charged in such a way so that gauge anomaly
is absent. In Table \ref{tab1} and \ref{tab2}, we have shown the SM particles and beyond SM
particles with the corresponding charges under the complete gauge group 
$SU(3)_{c} \times SU(2)_{L} \times U(1)_{Y} \times U(1)_{X}$.

\begin{center}
\begin{table}[h!]
\begin{tabular}{||c|c|c||}
\hline
\hline
\begin{tabular}{c}
    Gauge\\
    Group\\ 
    \hline
    
    ${\rm SU(3)}_{c}$\\ 
    \hline
    ${\rm SU(2)}_{\rm L}$\\ 
    \hline
    $U(1)_{Y}$\\ 
    \hline
    $U(1)_{X}$\\ 
\end{tabular}
&

\begin{tabular}{c|c|c|c|c|c|c|c|c}
    \multicolumn{9}{c}{Fermions}\\ 
    \hline
    $Q_i$&$U^c_i$&$D^c_i$&$L_{e}$&$L_{\mu}$&$L_{\tau}$&$E^c_{e}$&$E^c_{\mu}$&$E^c_{\tau}$\\ 
    \hline
    $3$&$\bar{3}$&$\bar{3}$&$1$&$1$&$1$&$1$&$1$&$1$\\ 
    \hline
    $2$&$1$&$1$&$2$&$2$&$2$&$1$&$1$&$1$\\ 
    \hline
    $1/6$&$-2/3$&$1/3$&$-1/2$&$-1/2$&$-1/2$&$1$&$1$&$1$\\
    \hline
    $0$&$0$&$0$&$-n$&$0$&$n$&$(n-1)$&$0$&$-(n-1)$\\ 
\end{tabular}
&
\begin{tabular}{c}
    \multicolumn{1}{c}{Scalars}\\
    \hline
    $\phi_h$\\
    \hline
    $1$\\
    \hline
    $2$\\
    \hline
     $-1/2$\\
    \hline
     $0$\\
\end{tabular}\\
\hline
\hline
\end{tabular}
\caption{SM particles and their corresponding charges under complete gauge group.}
\label{tab1}
\end{table}
\end{center}

\begin{center}
\begin{table}[h!]
\begin{tabular}{||c|c|c||}
\hline
\hline
\begin{tabular}{c}
    Gauge\\
    Group\\ 
    \hline
    
    ${\rm SU(3)}_{c}$\\ 
    \hline
    ${\rm SU(2)}_{\rm L}$\\ 
    \hline
    $U(1)_{Y}$\\ 
    \hline
    $U(1)_{L_{\tau} - L_{e}}$\\ 
\end{tabular}
&

\begin{tabular}{c|c|c|c|c|c|c}
    \multicolumn{7}{c}{Fermions}\\ 
    \hline
    $N^c_{e}$&$N^c_{\mu}$&$N^c_{\tau}$&$VL_{l1}$&$VL^c_{l1}$&$VL_{l3}$&$VL^c_{l3}$\\ 
    \hline
    $1$&$1$&$1$&$1$&$1$&$1$&$1$\\ 
    \hline
    $1$&$1$&$1$&$2$&$2$&$2$&$2$\\ 
    \hline
    $0$&$0$&$0$&$-1/2$&$1/2$&$-1/2$&$1/2$\\
    \hline
    $n$&$0$&$-n$&$-(n-1)$&$(n-1)$&$(n-1)$&$-(n-1)$\\ 
\end{tabular}
&
\begin{tabular}{c|c}
    \multicolumn{2}{c}{Scalars}\\
    \hline
    $\phi_1$ & $\phi_2$\\
    \hline
    $1$ & $1$\\
    \hline
    $1$ & $1$\\
    \hline
     $0$ & $0$\\
    \hline
     $1$ & $n$\\
\end{tabular}\\
\hline
\hline
\end{tabular}
\caption{BSM particles and their corresponding charges under complete gauge group.}
\label{tab2}
\end{table}
\end{center}

The complete Lagrangian for the above particle spectrum which consist of kinetic term, Yukawa 
term and potential is as follows,
\begin{eqnarray}
\mathcal{L} &=& \mathcal{L}_{kin} + \mathcal{L}_{lepton} + \mathcal{L}^{\phi_i}_{kin}
  + \mathcal{L}_{N}+
|D_{\mu} \phi_h|^2  + y^{u}_{ij} Q_{i} U^{c}_{j} \bar{H} 
+ y^{d}_{ij} Q_{i} D^{c}_{j} H  + y^{VL}_{1} VL_{l1} E^{c}_{1} H  \nn \\ 
&& + y^{VL}_{3} VL_{l3} E^{c}_{3} H  + y^{\phi_1}_{1} \phi_{1} L_{e} VL^{c}_{l1} 
+ y^{\phi_1}_{3} \bar{\phi_{1}} L_{\tau} VL^{c}_{l3} - M_{VL_{l1}} VL_{l1} VL^{c}_{l1}  
- M_{VL_{l3}} VL_{l3} VL^{c}_{l3}
\nn \\ && - \mathcal{V}(\phi_{h}, \phi_{1}, \phi_{2})
\label{lagrangian}   
\end{eqnarray}
where $\mathcal{L}_{kin}$ is the kinetic term for all the fermions and has the general form
$\mathcal{L}_{kin} = \bar{f} \gamma_{\mu} D_{\mu} f$, f is the corresponding
fermion and $D_{\mu}$ is the covariant
derivative with different form depending on the gauge charges of fermion $f$. $\mathcal{L}_{lep}$
contains the Yukawa terms associated with the leptons and further discussion on it is given in 
Eq.\,(\ref{lepton-lagrangian}) in Section \ref{xenon-1T-discussion}.
$\mathcal{L}^{\phi_i}_{kin}$ is the kinetic term for the extra singlet scalars $\phi_i$ (i = 1, 2) 
and $U(1)_{X}$ gauge boson $Z^{\prime}$ as shown in Eq.\,(\ref{kin-phi1-phi2}). 
$\mathcal{L}_{N}$ is the Lagrangian associated with the Dirac neutrino mass of the 
neutrinos and the Majorana mass term for the right handed neutrinos,
\begin{eqnarray}
 \mathcal{L}_{N} &=& y_{ee} L_{e} \phi_h N^c_{e} + y_{\mu \mu} L_{\mu} \phi_h N^c_{\mu} 
+ y_{\tau \tau} L_{\tau} \phi_h N^c_{\tau}
+ y^{\phi_2}_{e \mu} L_{e} \phi_h N^{c}_{\mu} \frac{\phi_{2}}{M_{Pl}} + 
y^{\phi_2}_{\mu e} L_{\mu} \phi_h N^{c}_{e} \frac{\phi^{\dagger}_{2}}{M_{Pl}} \nn \\
&& + y^{\phi_2}_{\mu \tau} L_{\mu} \phi_h N_{\tau} \frac{\phi_{2}}{M_{Pl}} +
 y^{\phi_2}_{\tau \mu} L_{\tau} \phi_{h} N^{c}_{\mu} \frac{\phi^{\dagger}_{2}}{M_{Pl}}  +
 Y_{e \mu} N^{c}_{e} N^{c}_{\mu} \phi^{\dagger}_{2} + M_{e \tau} N^{c}_{e} N^{c}_{\tau} 
 + M_{\mu \mu} N^{c}_{\mu} N^{c}_{\mu}  \nn \\ && 
 + Y_{\mu \tau} N^{c}_{\mu} N^{c}_{\tau} \phi_{2} + {\it h.c.} 
\label{neutrino-mass-matrix}
\end{eqnarray} 

The potential for the present model takes the following form,
\begin{eqnarray}
\mathcal{V}(\phi_h, \phi_1, \phi_2) &=& - \mu^{2}_{h} (\phi^{\dagger}_{h} \phi_h) 
+ \lambda_{h} (\phi^{\dagger}_{h} \phi_h)^2 - \mu^{2}_{i} (\phi^{\dagger}_{i} \phi_i)
+ \sum_{i = 1, 2} \lambda_{\phi_i} (\phi^{\dagger}_{i} \phi_i)^2 
+ \lambda_{12} (\phi^{\dagger}_{1} \phi_1) (\phi^{\dagger}_{2} \phi_2)  \nn \\ &&
+ \sum_{i = 1, 2} \lambda_{\phi_{h} \phi_{i}} (\phi^{\dagger}_{h} \phi_h) (\phi^{\dagger}_{i} \phi_i)
+ \left[ \frac{\lambda_{\phi_{1} \phi_{2}} \phi^{n}_{1} \phi^{\dagger}_{2}}{M^{n-3}_{Pl}} + 
{\it h.c.}\right]
\label{hdo-potential}
\end{eqnarray}
Scalars take the following form at the time of the symmetry breaking 
\begin{eqnarray}
\phi_{h} = \begin{pmatrix}
0  \\
\frac{H + v}{\sqrt{2}}
\end{pmatrix}\,,
\,\,\,
\phi_{1} = \left( \frac{H_1 + v_{1}}{\sqrt{2}} \right) e^{i \frac{a_{1}}{v_{1}}}
\,\,\,{\rm and} \,\,\, \phi_{2} = \left( \frac{H_2 + v_{2}}{\sqrt{2}} \right) e^{i \frac{a_{2}}{v_{2}}}\,.
\end{eqnarray}
In the above equation we have shown the Higgses after taking vevs; in particular
we have written the SM Higgs doublet in the Unitary gauge
and for the other two singlets with their CP odd components. Due to the higher dimensional operator
of the singlet Higgses at the Planck scale,
the CP odd components will mix among each other and one of them is absorbed by the extra gauge 
boson and the remaining one will act as keV scale ALP, which is also FIMP dark matter in the present model
as will be discussed in the later part of the manuscript. 
From the scalar potential, we can determine the tadpole free conditions and the Higgs's masses, which are as follows,
\begin{eqnarray}
&&\left( \frac{\partial \mathcal{V}(\phi_{h}, \phi_{1}, \phi_{2})}{\partial \phi_j } \right)_{v, v_{1}, v_{2}} = 0\,\,\, {\rm and}\,\,\, 
\left( M^{2}_{H} \right)_{ij} =  \left( \frac{\partial^2 \mathcal{V}(\phi_{h}, \phi_{1}, \phi_{2})}{\partial \phi_i \partial \phi_j } \right)_{v, v_{1}, v_{2}}\,,
\end{eqnarray} 
where $\phi_{i, j} = H, H_{1}, H_{2}$.
Since, the neutral part of the Higgs fields are not directly related with the phenomenology in the present work, 
we neglect further discussion about its diagonalisation and defining the mass eigenstates for the Higgses which are throughly studied in the literature,
 \begin{eqnarray}
\begin{pmatrix}
h \\
h_1 \\
h_2
\end{pmatrix} 
=
\begin{pmatrix}
c^{\prime}_{12} c^{\prime}_{13} & s^{\prime}_{12} c^{\prime}_{13} & s^{\prime}_{13} \\
-s^{\prime}_{12} c^{\prime}_{23} - c^{\prime}_{12} s^{\prime}_{23} s^{\prime}_{13} & c^{\prime}_{12} c^{\prime}_{23} - s^{\prime}_{12} s^{\prime}_{23} s^{\prime}_{13} & s^{\prime}_{23} s^{\prime}_{13} \\
s^{\prime}_{12} s^{\prime}_{23} - c^{\prime}_{12} c^{\prime}_{23} s^{\prime}_{13} & -c^{\prime}_{12} s^{\prime}_{23} - s^{\prime}_{12} c^{\prime}_{23} s^{\prime}_{13} & c^{\prime}_{23} c^{\prime}_{13}
\end{pmatrix}
\begin{pmatrix}
H \\
H_1 \\
H_2
\end{pmatrix}
\end{eqnarray}
where $c^{\prime}_{ij} = \cos \theta^{\prime}_{ij}$ and $s^{\prime}_{ij} = \sin \theta^{\prime}_{ij}$.

\section{Neutrino Mass}
\label{neutrino-mass}
As given in Eq.\,(\ref{neutrino-mass-matrix}), we can write down the neutrino mass matrix after the 
$SU(3)_{c} \times SU(2)_{L} \times U(1)_{Y} \times U(1)_{X}$
gauge group breaks down to $SU(3)_{c} \times U(1)_{em}$ as, 
\begin{eqnarray}
- \mathcal{L}_{N} = 
\begin{pmatrix}
\nu &
N^{c}
\end{pmatrix}
\begin{pmatrix}
0 & M^{T}_{D} \\
M_{D} & M_{R}
\end{pmatrix}
\begin{pmatrix}
\nu \\
N^{c}
\end{pmatrix}
\end{eqnarray}
where $\nu = (\nu_{e}, \nu_{\mu}, \nu_{\tau})$, $N^{c} = (N^{c}_{e}, N^{c}_{\mu}, N^{c}_{\tau})$.
The Dirac mass matrix ($M_{D}$) and the Majorana mass matrix ($M_{R}$) take the following form,
\begin{eqnarray}
M_{D} = m_{d\,ij} = \begin{pmatrix}
\frac{y_{ee} v}{\sqrt{2}} & \frac{y_{e\mu} v v_{2}}{\sqrt{2} M_{Pl}} & 0  \\
\frac{y_{\mu e} v v_{2}}{\sqrt{2} M_{Pl}} & \frac{y_{\mu \mu} v }{\sqrt{2}} & \frac{y_{\mu \tau} v v_{2}}{\sqrt{2} M_{Pl}}  \\
0 & \frac{y_{\tau \mu} v v_{2}}{\sqrt{2} M_{Pl}} & \frac{y_{\tau \tau} v}{\sqrt{2}}  
\end{pmatrix}\,,\,\,{i,j = e,\mu,\tau}
\end{eqnarray}

\begin{eqnarray}
M_{R} = m_{R\,ij} = \begin{pmatrix}
0 & \frac{Y_{e\mu}v_{2}}{\sqrt{2}} & M_{e \tau} e^{i \theta}  \\
\frac{Y_{e\mu}v_{2}}{\sqrt{2}} & M_{\mu \mu} & \frac{Y_{\mu \tau}v_{2}}{\sqrt{2}}  \\
M_{e \tau} e^{i \theta} & \frac{Y_{\mu \tau}v_{2}}{\sqrt{2}} & 0  
\end{pmatrix}\,,\,\,{i,j = e, \mu, \tau}\,.
\label{neutrino-mass-in-basis}
\end{eqnarray}
In the case of the Majorana mass matrix, without loss of generality
we have only considered Yukawa terms up to dimension four.
In the seesaw approximation after diagonalising the matrix 
(as shown in Eq.\,(\ref{neutrino-mass-in-basis})),
one can have the light neutrino masses and heavy neutrino masses in the following form, 
Type-I seesaw mechanisms,
\begin{eqnarray}
m_{\nu} &=& - M^{T}_{D} M^{-1}_{R} M_{D} \nn \\
M_{N} &=& M_{R}
\end{eqnarray}

To obtain the allowed parameter space, we have considered neutrino oscillation parameters namely
sum of the light neutrino masses ($\sum_{i = 1,2,3} m_{\nu_i}$), 
two mass square differences ($\Delta m^2_{12}$, $\Delta m^2_{13}$) and three mixing angles 
($\theta_{12}$, $\theta_{13}$ and $\theta_{23}$) in the 
following range as obtained from Planck collaboration \cite{Ade:2015xua} 
and neutrino oscillation experiments \cite{Capozzi:2016rtj},
\begin{itemize}
\item cosmological upper bound on the sum of all three light neutrinos,
$\sum_i m_{i} < 0.23$ eV at $2\sigma$ C.L. \cite{Ade:2015xua},
\item mass squared differences $6.93<\dfrac{\Delta m^2_{21}}
{10^{-5}}\,{\text{eV}^2} < 7.97$ and $2.37<\dfrac{\Delta m^2_{31}}
{10^{-3}}\,{\text{eV}^2} < 2.63$ in $3\sigma$ range \cite{Capozzi:2016rtj},
\item all three mixing angles $30^{\circ}<\,\theta_{12}\,<36.51^{\circ}$,
$37.99^{\circ}<\,\theta_{23}\,<51.71^{\circ}$ and
$7.82^{\circ}<\,\theta_{13}\,<9.02^{\circ}$ also in $3\sigma$ range
\cite{Capozzi:2016rtj}.
\end{itemize} 

In generating the scatter plots among the model parameters, we have diagonalised 
the neutrino mass matrices and collected those points which satisfy the 
neutrino oscillation data as given earlier. We have varied neutrino mass matrix elements in the following range,
\begin{eqnarray}
10^{-8}\,\,{\rm GeV} \leq & m_{d\,ij}& \leq 10^{-4}\,\,{\rm GeV} \nn \\
10^{-6}\,\,{\rm GeV} \leq & m_{R\,ij}& \leq 10^{2}\,\,{\rm GeV} \nn \\
0 \leq &\theta &\leq \pi
\end{eqnarray} 
where $i,j$ = e, $\mu$, $\tau$. Now, we are going to show few scatter plots which are 
obtained for the normal hierarchy of the neutrino mass matrix and one can obtain similar kind of plots
for the inverted hierarchy (IH) of the neutrino masses.
\begin{figure}[h!]
\centering
\includegraphics[angle=0,height=7.5cm,width=8.5cm]{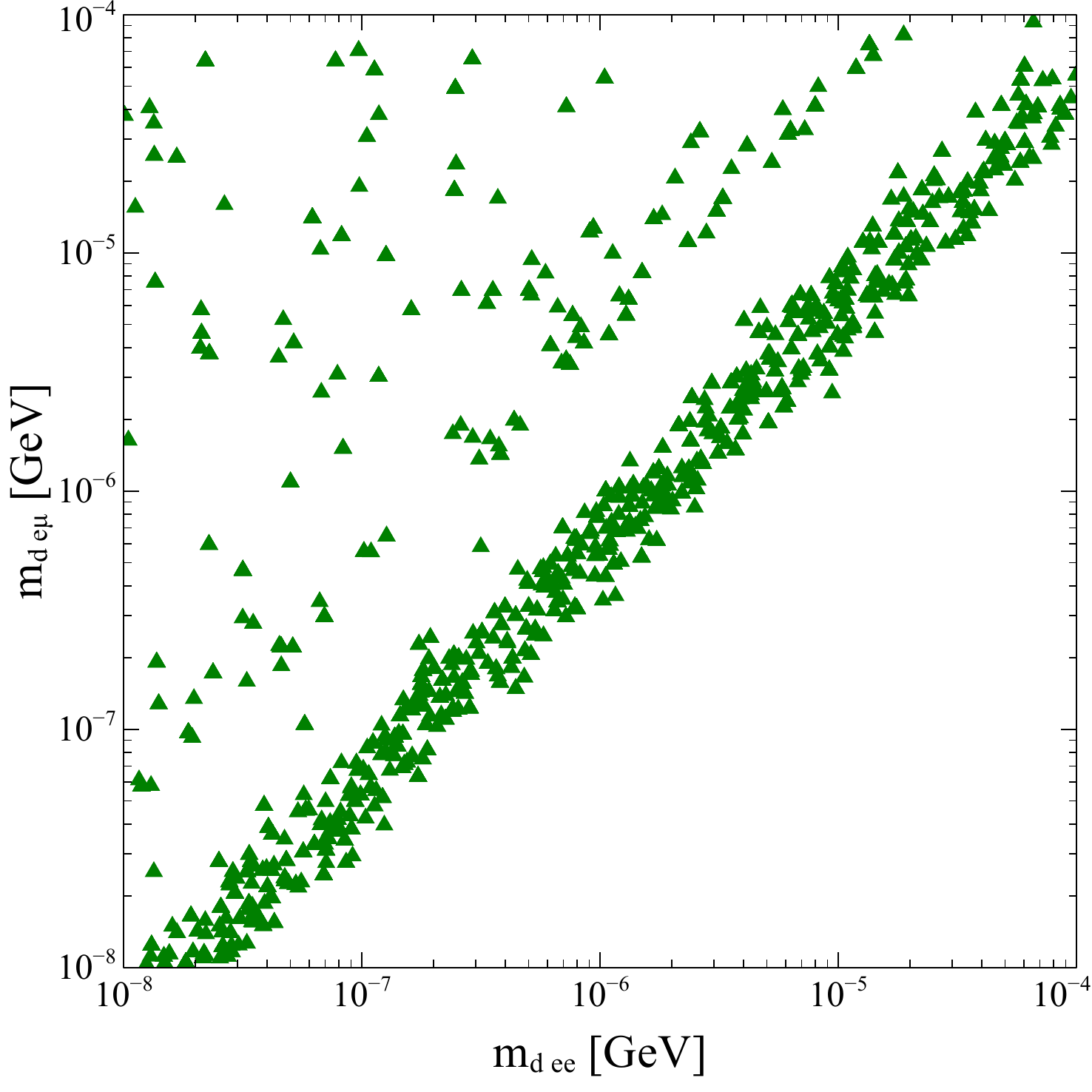}
\includegraphics[angle=0,height=7.5cm,width=8.5cm]{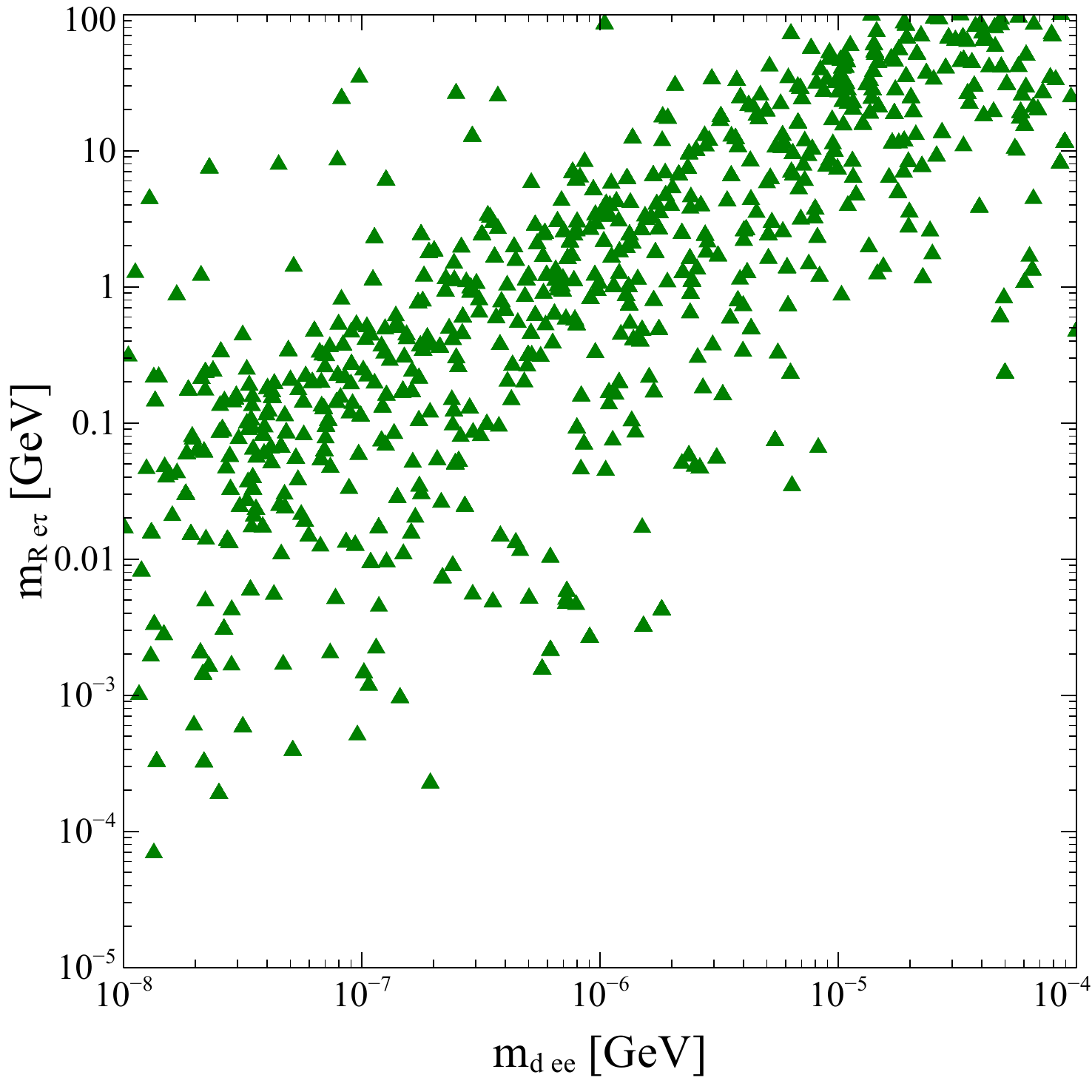}
\caption{LP (RP): Scatter plot in the $m_{d\,ee}-m_{d\,e\mu}$ ($m_{d\,ee}-m_{R\,e\tau}$) plane after 
satisfying neutrino oscillation data.}
\label{neutrino-fig-1}
\end{figure}

In the LP of Fig.\,(\ref{neutrino-fig-1}), we have shown the variation in the 
$m_{d\,ee}-m_{d\,e\mu}$ plane after satisfying the $3\sigma$ bound on two mass square 
differences and three mixing angles as listed before. One can easily see from the figure that
there exist a sharp correlation between the $m_{d\,ee}$ and $m_{d\,e\mu}$ parameters. They also
lie around the same ballpark value. One can see that most of the points satisfy the ratio
$\frac{m_{d\,e\mu}}{m_{d\,ee}} \sim 0.6$ and from oscillation experiments we also know that
$\tan(\theta_{12}) \sim 0.6$. So we can conclude that this type of sharp correlation is coming from the $\theta_{12}$ bound of the oscillation experiment. There are other points
also on the upper side due to the variation of the other parameters. On the other
hand in the RP of the figure, we have shown the variation in the $m_{d\,ee}-m_{R\,e\tau}$
plane after satisfying the oscillation data. In this figure, we can also notice that there exist a correlation between these two parameters. This correlation is trying to obey
the bound $\frac{m^2_{d\,ee}}{m_{R\,e\tau}} \leq 10^{-10}$ GeV which is coming from the 
two mass square differences $\Delta m^2_{12}$, $\Delta m^2_{13}$ and sum of neutrino masses
$\sum_{i=1,2,3} m_{\nu_i}$. There is also upper bound in the same plane for lower values of 
$m_{d\,ee}$ which is just coming from the bound $\frac{m^2_{d\,ee}}{m_{R\,e\tau}} \geq 10^{-15}$
because below this value mass the square differences are very difficult to obtain in the 
$3\sigma$ range.  
  
\begin{figure}[h!]
\centering
\includegraphics[angle=0,height=7.5cm,width=8.5cm]{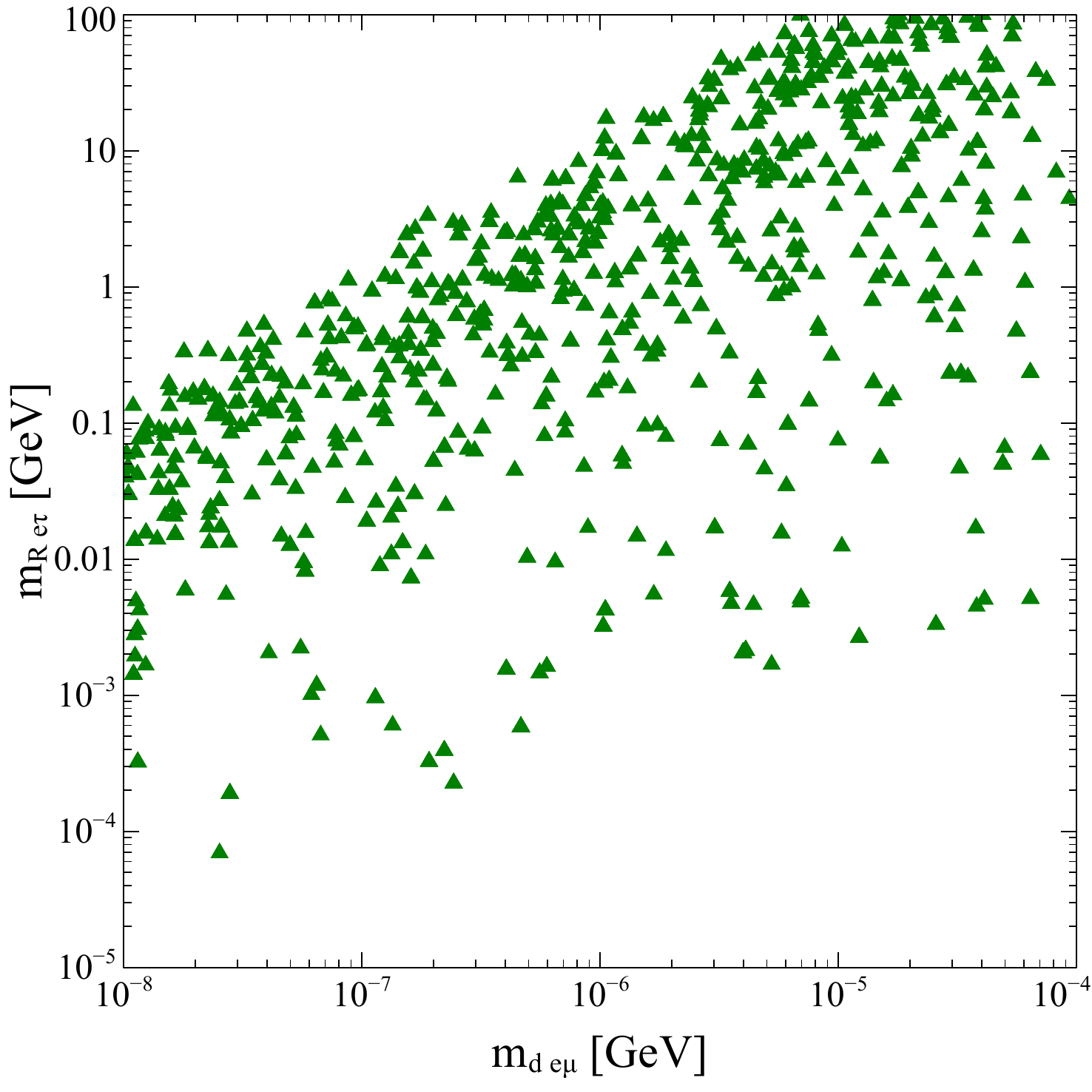}
\includegraphics[angle=0,height=7.5cm,width=8.5cm]{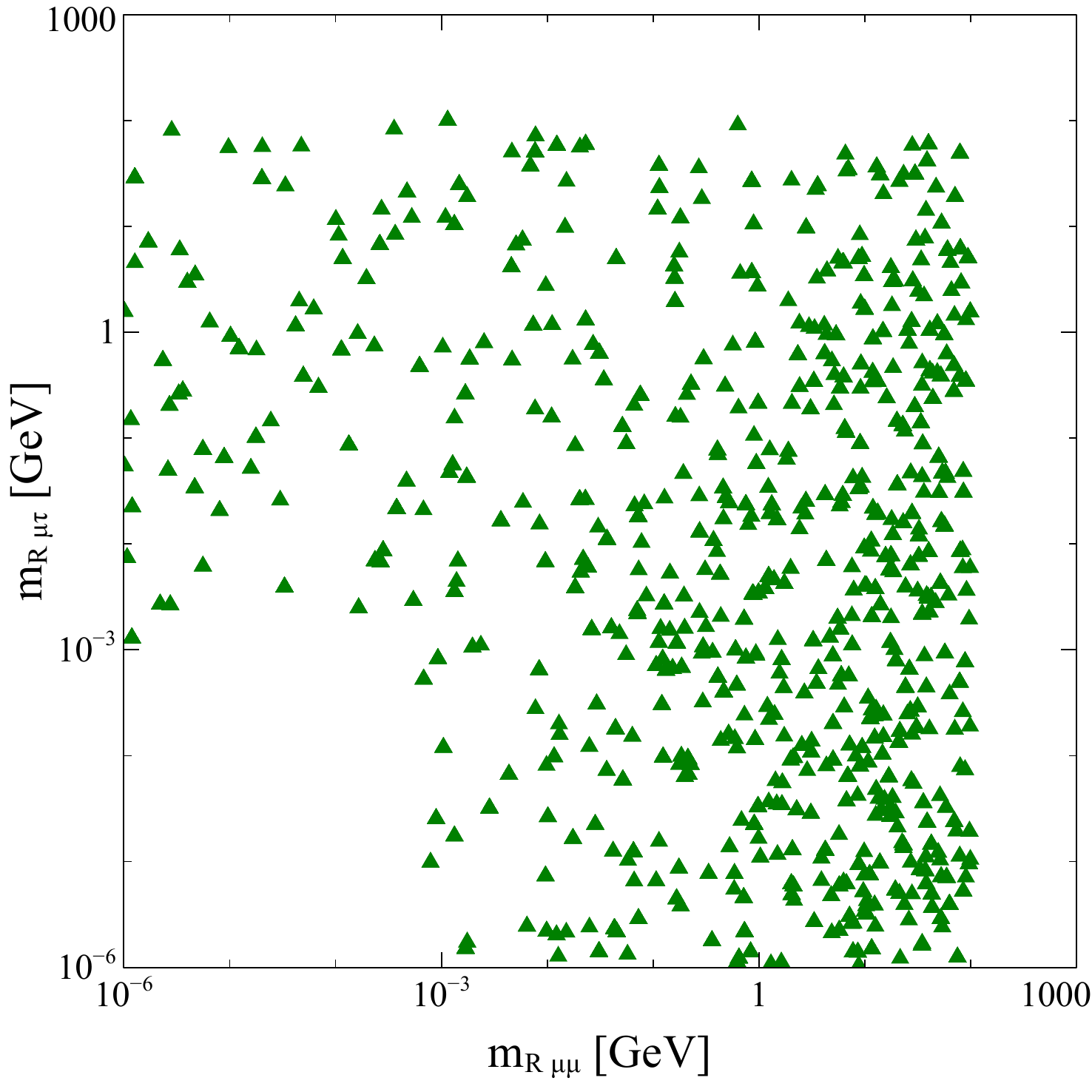}
\caption{LP (RP): Scatter plot in the $m_{d\,e\mu}-m_{R\,e\tau}$ ($m_{R\,\mu\mu}-m_{R\,\mu\tau}$) 
plane after satisfying neutrino oscillation data.}
\label{neutrino-fig-2}
\end{figure}

Fig.\,(\ref{neutrino-fig-2}) shows the variation in the
$m_{d\,e\mu}-m_{R\,e\tau}$ and $m_{R\,\mu\mu}-m_{R\,\mu\tau}$ planes after   
satisfying the neutrino oscillation data. In particular, the LP draws the same kind of physics conclusion
as explained in the RP of Fig.\,(\ref{neutrino-fig-1}). Here also the lower bound is coming 
from constraint $\frac{m^2_{d\,e\mu}}{m_{R\,e\tau}} \leq 10^{-10}$ and a sharp upper bound is coming 
from $\frac{m^2_{d\,e\mu}}{m_{R\,e\tau}} \geq 10^{-14}$. In the RP we have shown the variation
in $m_{R\,\mu\mu}-m_{R\,\mu\tau}$ plane after satisfying the $3\sigma$ bound.
We can see from the figure that both the parameter can not take low values simultaneously
but they can take higher values together and are not ruled out by the oscillation data. Moreover,
either of them can take a low value but at the same time, the other parameter
has to take a high value to satisfy the neutrino oscillation data.


\section{Xenon-1T signal}
\label{xenon-1T-discussion}
As said in the model part (Section \ref{model}), the CP even parts of the scalar fields are 
not directly related with 
the phenomenology we are interested in this work. On the other hand, the CP odd part of the singlet
Higgses take a pivotal role in the present work, hence we are going to discuss now the CP odd 
components. CP odd part can be written in the following manner,
\begin{eqnarray}
\phi_{1} = \frac{v_{1}}{\sqrt{2}} e^{i \frac{a_1}{v_1}}\,\,{\rm and}\,\, \phi_{2} = \frac{v_{2}}{\sqrt{2}} e^{i \frac{a_2}{v_2}}
\label{phi-1-phi-2-form}
\end{eqnarray} 
where it follows the {\it vev} condition $|\phi_i|^2 = \frac{v^{2}_{i}}{2}$. Kinetic term for the 
extra singlet scalars and ($U(1)_{X}$) gauge boson take the following form,
\begin{eqnarray}
\mathcal{L}^{\phi_i}_{kin} = -\frac{1}{4} F^{\prime}_{\mu\nu} F^{\prime\,\mu\nu}
+ \sum_{i=1,2} |D_{\mu} \phi_i|^2
\label{kin-phi1-phi2}
\end{eqnarray}
where $F^{\prime}_{\mu\nu}$ is the field strength tensor associated with 
$U(1)_{X}$ gauge group. The covariant derivative takes the following form,
\begin{eqnarray}
D^{i}_{\mu} = \partial_{\mu} - i g^{\prime} Z^{\prime} n_{i}
\end{eqnarray}
where $n_i$ and $g^{\prime}$ are the $U(1)_{X}$ gauge charge and gauge coupling.
By expanding the covariant derivative in Eq.\,(\ref{kin-phi1-phi2}), we have the kinetic term as
follows,
\begin{eqnarray}
\mathcal{L}^{\phi_i}_{kin} = -\frac{1}{4} \left( F^{\prime}_{\mu\nu} \right)^2 
+ \frac{1}{2} \left(\partial_{\mu} a_1 \right)^2 + \frac{1}{2} \left(\partial_{\mu} a_2 \right)^2
+ \frac{M^2_{Z^{\prime}}}{2} \left( Z^{\prime}_{\mu} \right)^2 - 
g^{\prime} Z^{\prime}_{\mu} \partial^{\mu} \left[n_{1} v_{1} a_{1} + n_{2} v_{2} a_{2} \right]
\label{modified-kinetic-term}
\end{eqnarray}
where the gauge boson mass has the following form,
\begin{eqnarray}
M_{Z^{\prime}} = g^{\prime \, 2} \left(n^2_{1} v^2_{1} + n^2_{2} v^2_{2} \right)\,.
\end{eqnarray}
We can redefine $a_1$ and $a_2$ in terms of mass basis $a$ and $G_{Z^{\prime}}$ as follows,
\begin{eqnarray}
a = \frac{1}{\sqrt{n^2_{1} v^2_{1} + n^2_{2} v^2_{2} }} 
\left[ n_{2} v_{2} a_{1} - n_{1} v_{1} a_{2} \right] \nn \\
G_{Z^{\prime}} = \frac{1}{\sqrt{n^2_{1} v^2_{1} + n^2_{2} v^2_{2} }} 
\left[ n_{1} v_{1} a_{1} + n_{2} v_{2} a_{2} \right]\,.
\end{eqnarray}
Therefore, in terms of $a$ and $G_{Z^{\prime}}$ Eq.\,(\ref{modified-kinetic-term}) takes the
following form,
\begin{eqnarray}
\mathcal{L}^{\phi_i}_{kin} = -\frac{1}{4} \left( F^{\prime}_{\mu\nu} \right)^2 + 
\frac{1}{2} \left(\partial_{\mu} a \right)^2 
+ \frac{1}{2} M^{2}_{Z^{\prime}} \left( Z^{\prime} 
- \frac{\partial_{\mu} G_{Z^{\prime}}}{M_{Z^{\prime}}} \right)^2
\end{eqnarray}
The above Lagrangian is in St\"{u}ckelberg form and it is clearly visible that the $G_{Z^{\prime}}$
degree of freedom is becoming the transverse component of $Z^{\prime}$ and makes it massive.
The other CP odd component ``$a$" is massless. If we consider the higher dimensional
operator which is consist of $\phi_1$ and $\phi_2$ as given in Eq.\,(\ref{hdo-potential}),
\begin{eqnarray}
\mathcal{V}^{HDO} = \frac{\lambda_{\phi_1 \phi_2}\phi^n_1 \phi^{\dagger}_2}{M^{n-3}_{Pl}} + {\it h.c.}\,.
\end{eqnarray}
 By using the form as given in Eq.\,(\ref{phi-1-phi-2-form}), we can write the above HDO in the 
 following way,
 \begin{eqnarray}
 \mathcal{V}^{HDO} &=& \frac{\lambda_{\phi_1 \phi_2} v^n_{1} v_{2}}{2^{\frac{n+1}{2}} M^{n-3}_{Pl}} 
 e^{i \left(\frac{n v_{2} a_{1} - v_{1} a_{2}}{v_{1} v_{2}} \right)} + {\it h.c.} \nn \\
 &=& \frac{\lambda_{\phi_1 \phi_2} v^n_{1} v_{2}}{2^{\frac{n+1}{2}} M^{n-3}_{Pl}} e^{i \frac{a}{V}} + {\it h.c.}
 \end{eqnarray}
where $V = \frac{v_{1} v_{2}}{\sqrt{n^2_{1} v^2_{1} + n^2_{2} v^2_{2} }}$ and we have used the
definition of $a$. Expanding the above expression we get,
\begin{eqnarray}
\mathcal{V}^{HDO} = \frac{2 \lambda_{\phi_1 \phi_2} v^n_{1} v_{2}}{2^{\frac{n+1}{2}} M^{n-3}_{Pl}}
 \left( 1 - \frac{a^{2}}{2 V^{2}} + ... \right)\,. 
\end{eqnarray}
Finally we get mass of ALP $a$ which is $m^2_{a} = \frac{2 \lambda_{\phi_1 \phi_2} v^n_{1} v_{2}}{2^{\frac{n+1}{2}} M^{n-3}_{Pl} V^{2}}$. 
\begin{figure}[h!]
\centering
\includegraphics[angle=0,height=7.5cm,width=8.5cm]{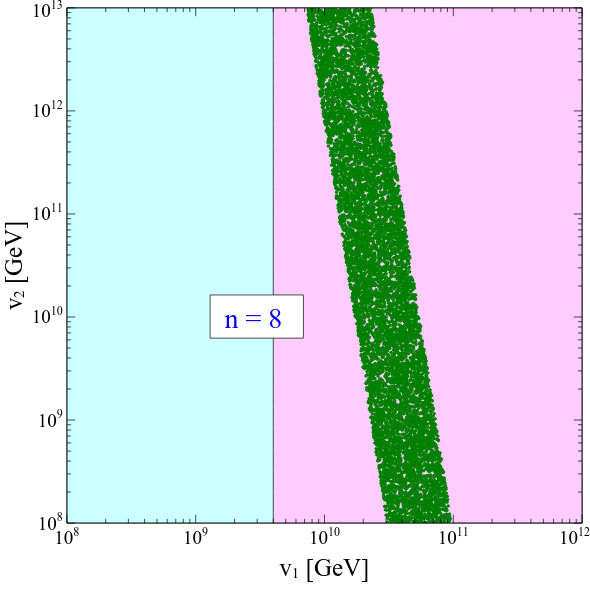}
\caption{Scatter plot in the $v_{1}-v_{2}$ plane where green
points are after demanding ALP mass in between 1-100 keV. Magenta 
region satisfies the electron and tauon mass by suitably choosing 
$y_{e}$, $y^{VL}_{1}$, $y^{VL}_{3}$ and $y_{\tau}$
values. Cyanide region is disallowed by the electron and tauon mass.
We have kept fixed $\lambda_{\phi_1 \phi_2}$ = 1, $y^{\phi_1}_{1,3} = 10^{-8}$ and considered reduced value of the Planck mass, $M_{Pl} = 2.43 \times 10^{18}$ GeV. }
\label{v1-v2-axion-mass}
\end{figure}

In Fig.\,(\ref{v1-v2-axion-mass}), we have shown the variation in the
$v_{1} - v_{2}$ plane for $n=8$ which is the $U(1)_{X}$ charge of $\phi_2$. 
Magenta region is allowed from the electron and tauon mass
whereas cyanide region is disallowed from the electron and muon mass. Green points are 
obtained after demanding the ALP mass in the 1-100 keV range. To satisfy, the electron mass, 
muon mass and axion mass in the keV range, we need $n\geq 8$ which is the $U(1)_{X}$
charge of $\phi_2$ singlet scalar. We can say from the electron, muon and keV order ALP
mass bound that the $U(1)_{X}$ charge of $n < 8$ is already ruled out. Although, in generating the 
scatter plot we have considered $\lambda_{\phi_1 \phi_2} = 1$ and if we take the other value of 
$\lambda_{\phi_1 \phi_2}$ then it will accordingly change the bound on $n$.        

Since our main target is to explain the Xenon-1T
signal, we have to discuss about the ALP coupling with electrons. As given in 
Eq.\,(\ref{lagrangian}) if we integrated out the vector like leptons 
(more discussion is in Section \ref{uv-contribution}) then we get the following terms,
\begin{eqnarray}
\mathcal{L}_{lepton} & \supset & 
-\frac{y^{VL}_{1} y^{\phi_1}_{1}}{M^{VL}_{1}} \phi_{1} L_{e} E^{c}_{e} \phi_{h} 
- y_{e} L_{e} \phi_h E^{c}_{e} \frac{\phi_{1}}{M_{Pl}} - y_{\mu} L_{\mu} \phi_h E^{c}_{\mu} 
-\frac{y^{VL}_{3} y^{\phi_1}_{3}}{M^{VL}_{3}} \bar{\phi_{1}} L_{\tau} E^{c}_{\tau} \phi_{h} 
\nn \\ && 
-y_{\tau} L_{\tau} \phi_h E^{c}_{\tau} \frac{\bar{\phi_{1}}}{M_{Pl}} + {\it h.c.}\,,
\label{hdo-freeze-in}
\end{eqnarray}
once $\phi_1$ and $\phi_h$ take vevs then we get,
\begin{eqnarray}
\mathcal{L}_{lepton} \supset -m_{e} l_{e} E^{c}_{e} e^{i \frac{a_{1}}{v_{1}}}
-m_{\mu} l_{\mu} E^{c}_{\mu} 
-m_{\tau} l_{\tau} E^{c}_{\tau} e^{-i \frac{a_{1}}{v_{1}}} + {\it h.c.}
\label{lepton-lagrangian}
\end{eqnarray}
where $m_{e} = \left( \frac{y^{VL}_{1} y^{\phi_1}_{1} v v_{1}}{2 M^{VL}_{1}} 
+ \frac{y_{e} v v_{1}}{2 M_{Pl}} \right)$, $m_{\mu} = \frac{y_{\mu} v}{\sqrt{2}}$ and 
$m_{\tau} = \left( \frac{y^{VL}_{3} y^{\phi_1}_{3} v v_{1}}{2 M^{VL}_{3}} 
+ 
\frac{y_{\tau} v v_{1}}{2 M_{Pl}} \right)$ are the electron, muon and tauon mass.
One important think to note here is that the ALP coupling to electron ($g_{aee}$)
is given by
\begin{eqnarray}
g_{aee} \simeq \frac{m_{e} V}{v^2_{1}} = 2.5 \times 10^{-14} 
\left( \frac{V \times 2 \times 10^{10}\,\,GeV }{v^2_{1}} \right)
\label{gaee-xenon1t}
\end{eqnarray}
We consider a global symmetry $U(1)_{g}$ and leptons are charged under this global symmetry.
Once the global symmetry gets broken then pseudo Nambu Goldstone boson (PNGB) is produced 
which acquire keV order mass from the higher dimensional operator as discussed before.
We assign the global charge in such a way so that it is anomaly free and naturally
the decay of PNGB to $\gamma \gamma$ is suppressed. We are working in keV range PNGB,
hence it will not decay directly to electrons but can decay to $\gamma \gamma$.
By integrating out electron and tauon, we generate the following kind of 
interaction \cite{Nakayama:2014cza},
\begin{eqnarray}
\mathcal{L}_{eff} \simeq -(1 - 1) \frac{\alpha_{em}}{4 \pi v_{1}} a_{1} F_{\mu\nu} \tilde{F}^{\mu\nu}
 + \frac{\alpha_{em}}{48 \pi v_{1}} \left( \frac{1}{m^2_{e}} - \frac{1}{m^2_{\tau}} \right)
 \left[ (\partial^{2} a_{1}) F_{\mu\nu} \tilde{F}^{\mu\nu} + 2 a_{1} F_{\mu\nu}\partial^2 \tilde{F}^{\mu\nu} \right]\,,
\end{eqnarray} 
first term comes from the anomaly $U(1)_{g}-U(1)_{em}-U(1)_{em}$ and we chose the global charge 
such that the anomaly is zero in the present case. The second term is the threshold correction. Since we will be dealing with the on shell PNGB, hence we can use the on shell condition
which turn the above equation into the following form,
\begin{eqnarray}
\mathcal{L}_{eff} \simeq \frac{\alpha_{em} V m^2_{a}}{48 \pi v^{2}_{1}} \left( \frac{1}{m^2_{e}} - \frac{1}{m^2_{\tau}} \right) a F_{\mu\nu} \tilde{F}^{\mu\nu}
\end{eqnarray}
where we have used $a_{1} = \frac{n_{1} v_{1} G_{Z^{\prime}} + n_{2} v_{2} a}
{\sqrt{n^{2}_{1} v^2_{1} + n^2_{2} v^2_{2}}}$\,. Decay rate of PNGB, a, to $\gamma\gamma$
is given by,
\begin{eqnarray}
\Gamma_{a \rightarrow \gamma\gamma} \simeq \frac{\alpha_{em} m^{7}_{a}}{9216 \pi^3 
(\frac{v^2_1}{V})^2} \times \frac{1}{m^4_{e}}\,.
\end{eqnarray}
We can now estimate the lifetime of the keV scale axion like particle a which only  
decay to $\gamma\gamma$ (since $m_{a} < 2 m_{l}$, $m_{l}$ is the lepton mass) is given by,
\begin{eqnarray}
\tau_{a \rightarrow \gamma \gamma} = \frac{1}{\Gamma_{a \rightarrow \gamma\gamma}}
\simeq 3.9 \times 10^{31} \left(\frac{m_{a}}{2.5\,\,keV}\right)^{-7} 
\left(\frac{v^2_{1}}{V\times 10^{10}\,\,GeV} \right)^{2} \,.
\end{eqnarray}
The decay lifetime of the ALP is larger than the age of the Universe and also safe from the 
X-ray bound, hence it can be a viable DM candidate. In the literature this kind of DM has been produced thermally and misalignment mechanism as described in \cite{Nakayama:2014cza, Li:2020naa}. By suitably choosing the 
ALP decay constant, one can reduce the thermal production of ALP. Moreover, We assume that inflation happens after the
global symmetry gets broken. Therefore, we can reduce the DM production by misalignment
mechanism by choosing the lower value of the  initial oscillation amplitude. 
In the present work, we are going to 
produce the keV scale DM by the freeze-in mechanism namely from the decay of the vector like 
fermion and from the HDO as given in Eqs.\,(\ref{lagrangian}), (\ref{hdo-freeze-in}). Now we are going to discuss the freeze-in production 
of keV scale ALP by two different kind of interactions.   

\subsection{keV scale FIMP DM}

\subsubsection{Decay Contribution}

\begin{figure}[h!]
\centering
\includegraphics[angle=0,height=7.5cm,width=8.5cm]{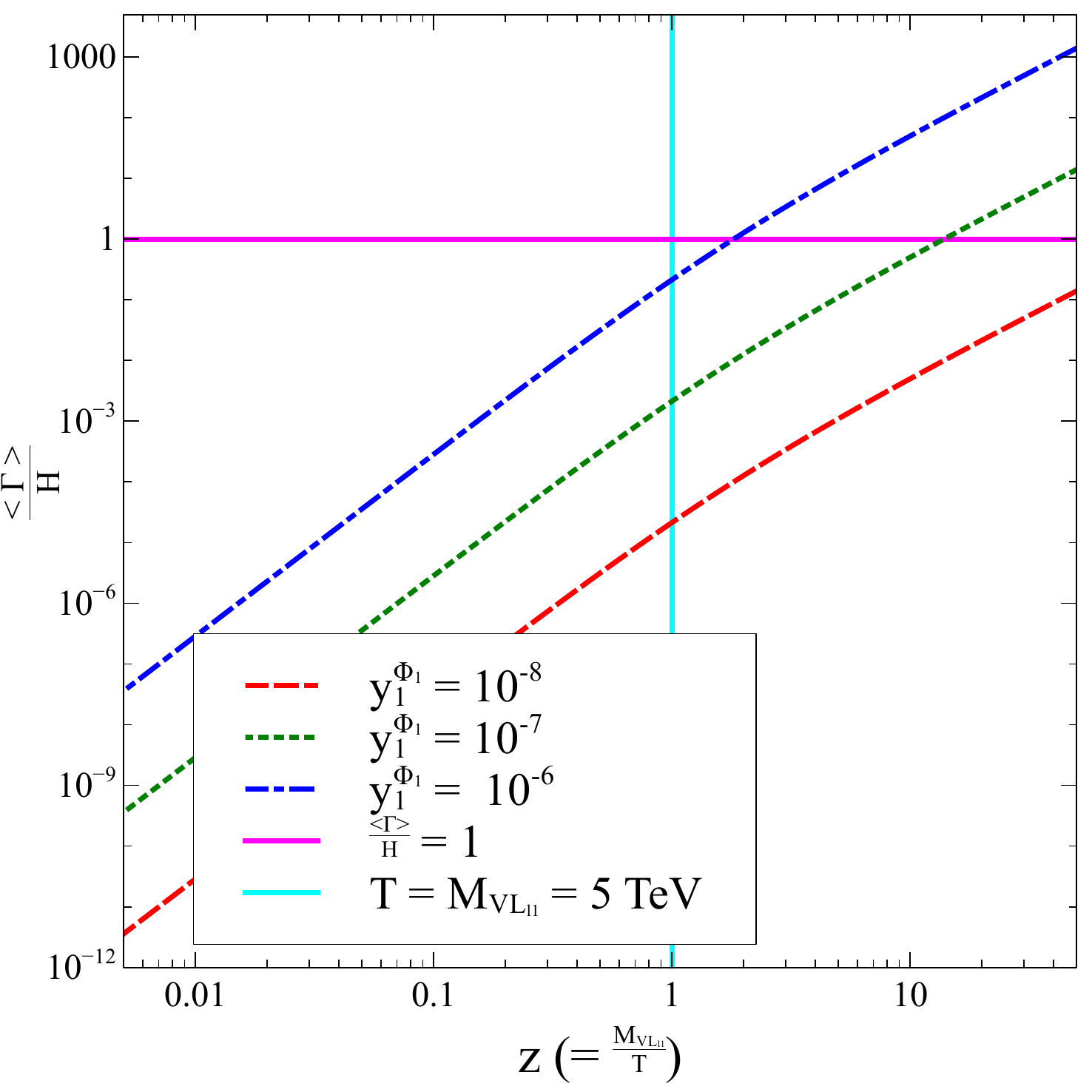}
\caption{$\frac{<\Gamma>}{H}$ variation with z for three different values of $y^{\phi_1}_{1}$.}
\label{gamma-by-h}
\end{figure}

As given in Eq.\,(\ref{lagrangian}), we can produce the keV scale DM from the decay of the vector like
doublet fermion through freeze-in mechanism by making the corresponding coupling in the feeble 
range. In Fig.\,(\ref{gamma-by-h}), we have shown the out of equilibrium condition of
keV range DM for 5 TeV vector like lepton for three different values of coupling strength.
In Fig.\,(\ref{gamma-by-h}), we have shown the variation of $\frac{<\Gamma>}{H}$ with $z$
where $<\Gamma> = \Gamma_{VL_{l1} \rightarrow L a} \frac{K_{1}(z)}{K_{2}(z)}$ and 
$H = \frac{1.66 \sqrt{g^{\rho}_{*}} T^{2}}{M_{Pl}}$ are the thermal average of the decay width and 
Hubble parameter, respectively. $K_{1}(z)$, $K_{2}(z)$ are the modified Bessel function for the first kind and second kind, $g^{\rho}_{*}$ is the matter d.o.f of the Universe and $z = \frac{M_{VL_{l1}}}{T}$. In the figure one can easily see that at $T \sim M_{VL_{l1}}$, the out of equilibrium condition
 $\frac{<\Gamma>}{H} < 1$ is always satisfied. If we increase the $y^{\phi_1}_1$ value greater than $10^{-6}$, then we say that 
it will reach equilibrium. So to be on the safe side we consider in the present work 
$y^{\phi_1}_{1,3} \sim 10^{-8}$ and the vector like lepton mass in the TeV range.   
The Boltzmann equation for determining the decay contribution takes the following form,
\begin{eqnarray}
\frac{d Y_{a}}{d z} = \frac{M_{Pl}}{1.66 M_{VL_{l1}}} \frac{z \sqrt{g_{*}(z)}}{g_{s}(z)}
\left[\sum_{i = 1, 3}  < \Gamma_{VL_{li} \rightarrow L a}  > \left( Y^{eq}_{VL_{li}} - Y_{a} \right) \right]
\end{eqnarray} 
where $Y_{a} = \frac{n_{a}}{s}$ is the comoving number density of a and s is the 
entropy, $s = \frac{2 \pi^2 g^s_{*} T^{3}}{45}$, of the Universe. $g_{*}(z)$ is a parameter
which depends on the matter ($g_{\rho}(z)$) and entropy ($g_{s}(z)$) d.o.f of the Universe
in the following way,
\begin{eqnarray}
\sqrt{g_{*}(z)} = \frac{g_{s}(z)}{\sqrt{g_{\rho}(z)}} \left( 1 - \frac{1}{3} 
\frac{d\, {\rm ln}\, g_{s}(z)}{d\, {\rm ln}\, z} \right)\,.
\end{eqnarray} 
$< \Gamma_{VL_{li} \rightarrow L a}  >$ is the thermal average of the decay width
of vector like lepton $VL_{li}$ (i=1, 3) as defined earlier.
 As shown in \cite{McDonald:2001vt, Hall:2009bx},
one can approximately solve the above Boltzmann equation and gets the following analytical 
expression of the DM relic density,
\begin{eqnarray}
\Omega_{a} h^{2} = \sum_{i = 1, 3} \frac{1.09 \times 10^{27} g_{VL_{li}} m_{a} \Gamma_{VL_{li}}}
{g^{s}_{*} \sqrt{g^{\rho}_{*}} M^2_{VL_{li}}}
\label{relic-density-decay}
\end{eqnarray}
\begin{figure}[h!]
\centering
\includegraphics[angle=0,height=7.5cm,width=8.5cm]{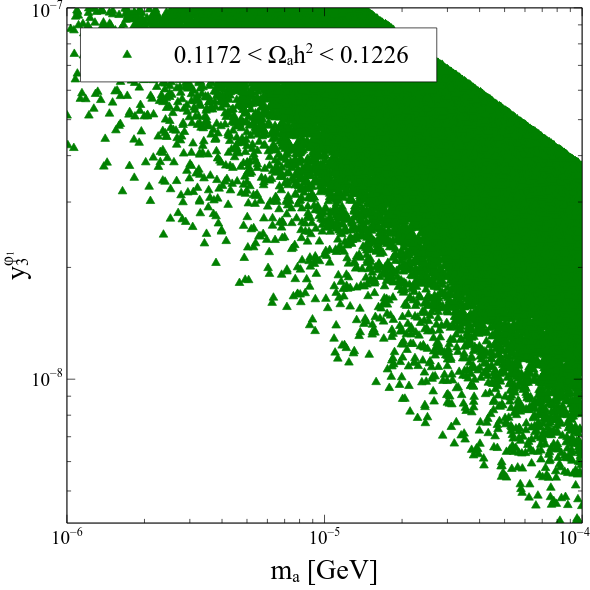}
\includegraphics[angle=0,height=7.5cm,width=8.5cm]{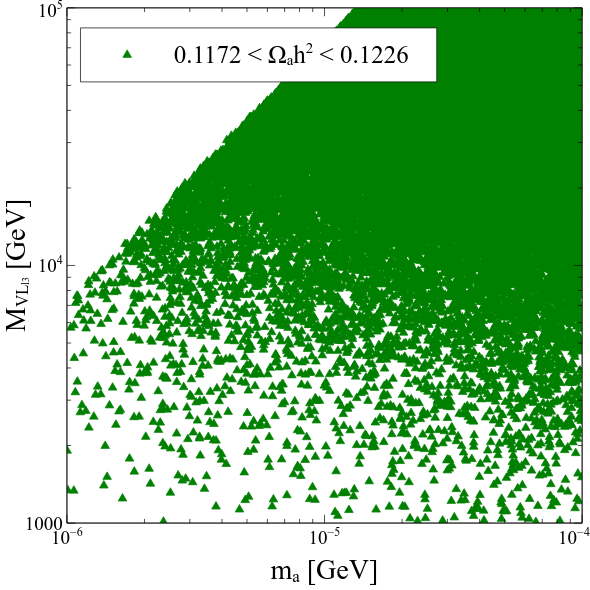}
\caption{LP (RP): Scatter plot in $m_{a} - y^{\phi_1}_{3}$ ($m_{a} - M_{VL_{l3}}$) plane after satisfying the DM relic density bound obtained from Planck satellite.}
\label{scatter-plot-decay}
\end{figure}
In generating the Fig.\,(\ref{scatter-plot-decay}), we have varied the three parameters
namely ALP mass $m_{a}$, Yukawa coupling $y^{\phi_1}_{3}$ and vector like lepton mass
$M_{VL_{l3}}$ in the following range \footnote{We have also assumed that same kind of contribution 
comes for $VL_{l1}$ decays and has been taken into account in the decay contribution.},
\begin{eqnarray}
1\,\,{\rm keV} \leq m_{a} \leq 100\,\,{\rm keV} \nn \\
10^{-7} \leq y^{\phi_1}_{3} \leq 10^{-10} \nn \\
1 \,\,{\rm TeV} \leq M_{VL_{l3}} \leq 100 \,\,{\rm TeV}\,
\end{eqnarray}
and we have used Eq.\,(\ref{relic-density-decay}) for calculating the DM relic density.
In the LP of Fig.\,(\ref{scatter-plot-decay}), we have shown the variation of dark matter mass
with the coupling $y^{\phi_1}_{3}$. Here, we have considered
vector like lepton $VL_{l3}$ decay to dark matter. All the points satisfy the DM relic density bound as obtained by the 
Planck collaboration \cite{Ade:2015xua}.
As given in Eq.\,{\ref{relic-density-decay}}, the decay width takes the form,
when the daughter particles have negligible mass in compared to mother particle,
\begin{eqnarray}
\Gamma_{VL_{l3}} = \frac{(y^{\phi_1}_{3})^{2}}{16 \pi} M_{VL_{l3}}\,.
\label{decay-width-VL}
\end{eqnarray}  
We can now easily see from Eq.\,(\ref{relic-density-decay}) that relic density is proportional to
the DM mass $m_{a}$ and the square of the Yukawa coupling $y^{\phi_1}_{3}$ 
{\it i.e.} $\Omega_{a} h^{2} \propto m_{a} (y^{\phi_1}_{3})^{2}$. To satisfy
the DM relic density it is clear that both of them can not increase or decrease 
simultaneously, in other words, there must exist anti-correlation between $m_{a}$ and $y^{\phi_1}_{3}$
which is visible in the LP of the figure. One can also notice that there is also 
a disallowed region in the upper corner of the figure which corresponds to the higher value of DM mass
and large Yukawa coupling. This region overproduces the DM hence is ruled out by the relic density bound.  
On the other hand in the RP of the figure, we have shown variation in the $m_{a} - M_{VL_{l3}}$
plane after satisfying the DM relic density. As can be seen from the 
Eq.\,(\ref{relic-density-decay}) the DM relic density varies as 
$\Omega_{a} h^{2} \propto \frac{m_{a}}{M_{VL_{l3}}}$. Therefore, to satisfy the DM relic density
bound from Planck we expect that both of them will either increase or decrease at the same time
which means we expect a sharp correlation among the parameters which is clearly shown in 
the RP of the figure.

\subsubsection{Contribution from higher dimensional operator}
\label{uv-contribution}

\begin{figure}[h!]
\centering
\includegraphics[angle=0,height=7.5cm,width=8.5cm]{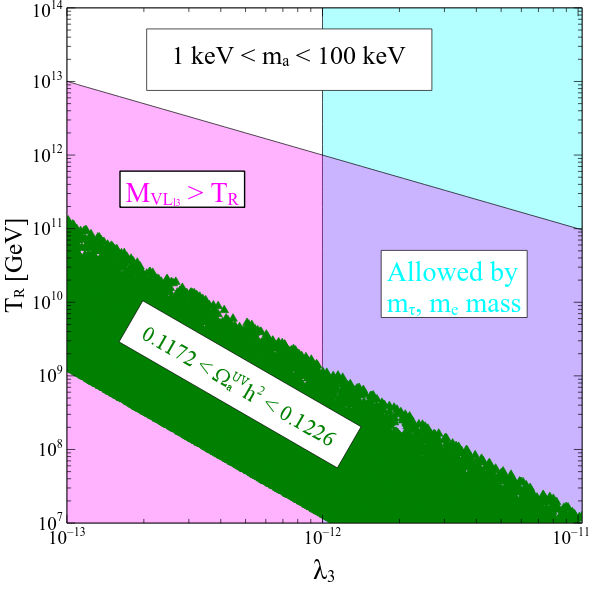}
\caption{Variation of reheat temperature ($T_{R}$) with the coupling $\lambda_{3}$ 
for DM mass between 1 to 100 keV. Magenta region corresponds to $M_{VL_{l3}} > T_{R}$ when
$y^{VL}_{3}, y^{\phi_1}_{3} \sim 1$. Cyan region is allowed by the electron and tauon mass. 
}
\label{relic-density-UV}
\end{figure}
Once we get rid of the vector type lepton $VL_{L1}$, $VL_{L3}$, then we get the following
dimension-5 operator,
\begin{eqnarray}
\mathcal{L}_{dim-5} = \frac{y^{VL}_{1} y^{\phi_1}_{1}}{M_{VL_{l1}}} \phi_1 L_{e} 
E^{c}_{e} \phi_{h} + \frac{y^{VL}_{3} y^{\phi_1}_{3}}{M_{VL_{l3}}} \bar{\phi_1} L_{\tau} 
E^{c}_{\tau} \phi_{h} + {\it h.c.} \,.
\label{hdo-me-mtau}
\end{eqnarray}   
As discussed in \cite{Hall:2009bx}, we can apply the freeze-in mechanism to produce the DM from the above 
non-renormalisable interaction terms. The coupling strength of the above interactions are
$\lambda_{i} = \frac{y^{VL}_{i} y^{\phi_1}_{i}}{M_{VL_{li}}} \sim 10^{-11}$ (i = 1, 3) which is in the 
ballpark value of the freeze-in coupling strength\footnote{$\lambda_1$ is suppressed by the electron
mass so in our case $\lambda_3$ takes an important role in DM production.}.
As discussed for decay contribution, vector like lepton is in thermal equilibrium
and ALP is produced from its decay. In this case to achieve tauon mass ($m_{\tau}$)
we need $v_{1} \sim 10^{16}$ GeV (where tauon mass, $m_{\tau} = \frac{y_{\tau} v v_1}{M_{Pl}}$) 
but this higher value of the vev can not
explain the Xenon-1T signal because it significantly reduces the ALP coupling strength
to electrons as given in Eq.\,(\ref{gaee-xenon1t}).
Therefore, to get the electron and tauon mass and the Xenon-1T benchmark point,
we have to consider the higher dimensional operator (HDO) as given in Eq.\,(\ref{hdo-me-mtau}).
Since the vector like lepton is integrated out, so in this case the vector like 
lepton mass ($M_{VL_{li}}$, i = 1, 3) has to be higher than the 
reheat temperature of the Universe. So, for this scenario when $M_{VL_{li}} > T_{R}$ (i =1, 3),
the ALP is not produced from the decay of $VL_{li}$ (i = 1, 3) but it is produced from the
HDO as given in Eq.\,(\ref{hdo-me-mtau}). For this case, the coupling $\lambda_3$ is suppressed
by the higher mass value of the vector like lepton and we can choose 
$y^{VL}_{3}, y^{\phi_1}_{3} \sim \mathcal{O}(1)$. 
  
In the present case, we will be considering higher values of vector like lepton mass and the reheat temperature 
(although $M_{VL_{l3}} > T_{R}$) which implies that
UV contribution is more relevant to us than the IR contribution. 
Therefore, we will be focusing on
the contribution coming from the non-renormalisable operator 
(as shown in Eq.\,(\ref{hdo-freeze-in}))
which depends on the unknown UV physics like reheat temperature $T_{R}$. Considering the 
non-renormalisable operator only, we can write down the Boltzmann equation for the production
of the ALP $a$ as follows \cite{Hall:2009bx},
     \begin{eqnarray}
     \frac{d n_{a}}{d t} + 3 n_{a} H \simeq \int d \pi_{a} d \phi_h d \pi_{L_{\tau}} d_{E^c_{\tau}}
     (2\pi)^4 \delta^{4}(p_{L_{\tau}} + p_{E^c_{\tau}} - p_{a} - p_{\phi_h})
      |M|^{2}_{L_{\tau} E^c_{\tau} \rightarrow a \phi_h} f_{L_{\tau}} f_{E^c_{\tau}}\,.
     \end{eqnarray}
After manipulating the above equation we get,
\begin{eqnarray}
\frac{d n_{a}}{d t} + 3 n_{a} H \simeq \frac{T}{2048 \pi^6}  \int ds d\Omega \sqrt{s}
      |M|^{2}_{L_{\tau} E^c_{\tau} \rightarrow a \phi_h} K_{1}\left(\frac{\sqrt{s}}{T}\right)\,,
\label{modified-BE}
\end{eqnarray}
where $s$ is the centre of mass energy of the $2 \rightarrow 2$ process.
Considering the fact that masses of the interacting particles are negligible compared to the
temperature we are working on. In this limit, the matrix element is expressed as 
$|M|^{2}_{L_{\tau} E^c_{\tau} \rightarrow a \phi_h} =  \lambda^2_{3}\, s$. After using this 
expression Eq.\,(\ref{modified-BE}) takes the following form,
\begin{eqnarray}
\frac{d n_{a}}{d t} + 3 n_{a} H \simeq  \frac{T \lambda^{2}_{3}}{512 \pi^5}
\int_{0}^{\infty} ds\, s^{3/2}\, K_{1}\left(\frac{\sqrt{s}}{T}\right)  
\end{eqnarray}       
Defining the comoving number density, $Y_{UV} = \frac{n_{a}}{s}$ and after integration, we get
\begin{eqnarray}
\frac{d Y_{UV}}{d T} \simeq - \frac{1 }{s H T} \frac{T^{6} \lambda^2_{3}}{16 \pi^5}\,.
\end{eqnarray}
Now, using the expression of entropy (s) and Hubble parameter (H), we get from the above equation 
after integration,
\begin{eqnarray}
Y_{UV} \simeq \frac{0.4 T_{R} \lambda^2_{3} M_{Pl}}
{\pi^{7} g^{s}_{*} \sqrt{g^{\rho}_{*}}}\,.
\label{Y-UV}
\end{eqnarray}   
Therefore, the relic density would be,
\begin{eqnarray}
\Omega^{UV}_{a} h^{2} \simeq 2.755 \times 10^{2} \left(\frac{m_{a}}{\rm keV}\right) Y_{UV}
\end{eqnarray}

In Fig.\,(\ref{relic-density-UV}), we have shown the allowed region in the 
$\lambda_{3} -T_{R}$ plane where  $\lambda_{3} = \frac{y^{VL}_{3} y^{\phi_1}_{3}}{M_{VL_{l3}}}$
and $T_{R}$ is the reheat temperature. 
Magenta region is coming when we impose the condition $M_{VL_{l3}} > T_{R}$ for $y^{VL}_{3}, y^{\phi}_{3} \sim 1$, 
whereas the cyan region is allowed by the electron and tauon mass.
All the green points satisfy the dark matter relic density put by
Planck collaboration. As seen in Eq.\,(\ref{Y-UV}), the comoving number density of the DM
varies as $Y_{UV} \propto T_{R} \lambda^2_{3}$ (as discussed the coupling 
$\lambda_{1}$ is suppressed due to electron mass so we are neglecting that coupling here). Therefore, the relic density as well as the 
comoving number density response to $\lambda_3$ and $T_R$ parameters anti-correlated way {\it i.e.}
DM relic density can be satisfied only when if $\lambda_3$ is increased then $T_{R}$ has to be decreased and vice versa. This type of behaviour is visible in the figure. The band of the green patch just indicates that there is also variation in the ALP mass in the range (1 - 100) keV. 
Finally, we say that the UV contribution can explain
the electron mass, tauon mass, Xenon-1T signal and DM relic density altogether.

\section{conclusion}
\label{conclusion}
In this work, we have considered a $U(1)_{X}$ gauge extension of the SM gauge group, 
leptons and extra particles are charged and quarks are neutral under this gauge group.
We have also extended the particle content by three right handed neutrinos, two vector like
leptons and two singlet scalars. We have assigned the $U(1)_{X}$ as well SM gauge group charges
to the particles in such a
way that the model is gauge anomaly free. Due to the presence of the right handed neutrinos, we can generate the
neutrino mass by the type-I seesaw mechanism. We have shown scatter plots among the neutrino mass
parameters after satisfying the neutrino oscillation data in $3\sigma$ range for normal hierarchy and one
can extrapolate this part for the inverted hierarchy as well. We also have two singlet scalars
and among the four d.o.f, two of them act as the physical Higgses, one is absorbed by
the gauge boson and the remaining one act like ALP. We can generate the coupling of the ALP
with electron and by suitably adjusting the parameter we can explain the Xenon-1T signal as well.
The global symmetry introduced is anomaly free, hence the axion coupling to photons is suppressed
and can evade the CXB easily. We have considered ALP as the FIMP type DM candidate
and has been produced from the decay of the vector like lepton. Moreover, ALP can be produced from the
higher dimensional operator as well by suitably adjusting the reheat temperature for heavy vector like lepton.
We have also pointed out that if we consider decay contribution to ALP production then it is difficult
to explain the lepton mass and Xenon-1T signal together. This problem can be resolved if we consider
the ALP production from the higher dimensional operator which appear when we integrated out the vector like lepton.
Since we have integrated out the vector like lepton, we have always followed that
vector like lepton mass is greater than the reheat temperature of the Universe. 
The present model can explain the neutrino mass, Xenon-1T signal through ALP interaction with the
electron. ALP can also serve as the viable DM candidate of the Universe which can be successfully produced by the freeze in mechanism either from the decay process or the annihilation
process coming from higher dimensional operator.

\section{Acknowledgements}
SK acknowledges the cluster computing facility at GWDG, G\"{o}ttingen, Germany.
SK is grateful to Laura Covi for revising the manuscript and discussion on the reheat 
temperature associated with the mass of vector like lepton.

\end{document}